\documentclass[twocolumn,aps,nofootinbib,twoside,superscriptaddress,showpacs,showkeys,preprintnumbers,amsmath,amssymb]{revtex4}
\usepackage{graphicx}

\def\acknowledgments{\section*{Acknowledgments}}%

\usepackage{amsthm}
\theoremstyle{definition}

\usepackage{bbm}

\def\R{{\mathbbm R}}

\def\SO{{SO}}

\def\etc{{\sl etc.\/}}

\let\phi=\varphi
\let\theta=\vartheta
\let\epsilon=\varepsilon

\def\tr{\mathop{\rm tr}\nolimits}

\newcommand{\color}[2][c]{}

\makeatletter

\newfont{\@aidxte}{cmsy10}
\newfont{\@aidxel}{cmsy10 scaled 1095}
\newfont{\@aidxtw}{cmsy10 scaled 1200}
\newlength\@aidxtexvi
\newlength\@aidxtexvii
\newlength\@aidxelxvi
\newlength\@aidxelxvii
\newlength\@aidxtwxvi
\newlength\@aidxtwxvii
\newcommand{\alignidx}[1]{%
  \@aidxtexvi=\fontdimen16\@aidxte
  \@aidxtexvii=\fontdimen17\@aidxte
  \@aidxelxvi=\fontdimen16\@aidxel
  \@aidxelxvii=\fontdimen17\@aidxel
  \@aidxtwxvi=\fontdimen16\@aidxtw
  \@aidxtwxvii=\fontdimen17\@aidxtw
    {\mbox{$%
    \fontdimen16\@aidxte=2.9pt
    \fontdimen17\@aidxte=2.9pt
    \fontdimen16\@aidxel=3.1pt
    \fontdimen17\@aidxel=3.1pt
    \fontdimen16\@aidxtw=3.3pt
    \fontdimen17\@aidxtw=3.3pt
    #1$}}%
    \fontdimen16\@aidxte=\@aidxtexvi
    \fontdimen17\@aidxte=\@aidxtexvii
    \fontdimen16\@aidxel=\@aidxelxvi
    \fontdimen17\@aidxel=\@aidxelxvii
    \fontdimen16\@aidxtw=\@aidxtwxvi
    \fontdimen17\@aidxtw=\@aidxtwxvii}

\makeatother

\def\del{\partial}
\def\emph#1{{\sl #1\/}}

\def\bra#1{\bigl<#1\bigr|}
\def\ket#1{\bigl|#1\bigr>}

\def\href#1{}
\def\dhref#1#2{}

\begin{document}
%

\preprint{DAMTP-2003-132}

\title{Diffeomorphisms from finite triangulations and absence\\
       of 'local' degrees of freedom}

\author{Hendryk Pfeiffer}
\email{H.Pfeiffer@damtp.cam.ac.uk}
\affiliation{Emmanuel College, St Andrew's Street, Cambridge CB2 3AP, United Kingdom}
\affiliation{DAMTP, Wilberforce Road, Cambridge CB3 0WA, United Kingdom}

\date{26 April 2004}

%
\begin{abstract}
%

If the diffeomorphism symmetry of general relativity is fully
implemented into a path integral quantum theory, the path integral
leads to a partition function which is an invariant of smooth
manifolds. We comment on the physical implications of results on the
classification of smooth and piecewise-linear $4$-manifolds which show
that the partition function can already be computed from a
triangulation of space-time. Such a triangulation characterizes the
topology and the differentiable structure, but is completely unrelated
to any physical cut-off. It can be arbitrarily refined without
affecting the physical predictions and without increasing the number
of degrees of freedom proportionally to the volume. Only refinements
at the boundary have a physical significance as long as the
experimenters who observe through this boundary, can increase the
resolution of their measurements. All these are consequences of the
symmetries. The Planck scale cut-off expected in quantum gravity is
rather a dynamical effect.

\end{abstract}


\maketitle 

%
\section{Introduction}
%

The quantum theory of gravity is expected to enjoy very special
properties some of which are tentatively stated by the
\emph{holographic principle}~\cite{Ho93}. For more details, we refer
to~\cite{Bo02}.

\textsl{(A) Holographic principle. In quantum gravity, classical
Lorentzian space-time with matter should emerge in such a way that the
number of independent quantum states associated with the light sheets
of any surface, is bounded by the exponential of the surface area.}

Associating the independent degrees of freedom with $2$-surfaces
rather than $3$-volumes is in outright contrast to the familiar
properties of local quantum field theory in a fixed metric
background. The following statement~\cite{Ho93,Bo02} can therefore
often be found in conjunction with~(A).

\textsl{(B) Breakdown of local quantum field theory (QFT). QFT whose
degrees of freedom are associated with the points of space, vastly
overestimates the number of degrees of freedom of the true quantum
theory of gravity.}

Principle~(A) essentially implies~(B). In this letter, we explain
how~(B) can already be resolved independently of~(A) if the
diffeomorphism symmetry of general relativity is implemented into the
path integral quantum theory. In order to resolve~(B) alone, no new
physical postulate is needed.

We first recall that any path integral quantization yields a partition
function that is an invariant of smooth manifolds. Results on the
classification of smooth and piecewise-linear (PL) manifolds show that
it can be computed from a triangulation of space-time which
characterizes the topology and the differentiable structure.

Naively, one would assume that the degrees of freedom were associated
with the simplices of the triangulation, that their number increased
with the number of simplices and that the metric size of the simplices
provided a short-distance cut-off. Theorems on the classification of
PL-manifolds, however, imply that the triangulation in the bulk can be
arbitrarily refined without affecting any physical prediction. The
actual degrees of freedom are therefore already determined by a
suitable \emph{minimal} triangulation. Only refinements at the
boundary are physically relevant as long as the experimenters who make
their observations through this boundary, can increase the resolution
of their measurements.

Whereas the precise form of the path integral for quantum gravity in
$d=3+1$ remains elusive, the spin foam models in $d=2+1$ precisely
implement the framework developed here. All properties described so
far, already follow from the symmetries and are independent of the
particular choice of variables or of the action.

%
\section{Path integrals}
%

\begin{figure}[t]
\begin{center}
\input{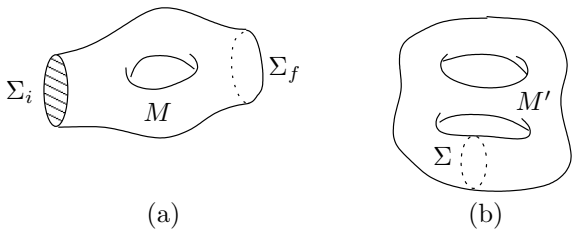}
\end{center}
\caption{
\label{fig_manifolds}
  (a) A four-manifold $M$ with boundary $\del M=\Sigma_i\cup\Sigma_f$.
  (b) If $\Sigma_i$ and $\Sigma_f$ are diffeomorphic, they can be
  identified to yield the closed manifold $M^\prime$.
}
\end{figure}

For simplicity, we consider general relativity without
matter. Space-time is a smooth oriented $4$-manifold with boundary
$\del M$. The classical theory~\cite{Wa84} is the study of the
existence and uniqueness of (smooth) metric tensors $g$ on $M$ that
satisfy the Einstein equations subject to suitable boundary
conditions. In the first order Hilbert--Palatini formulation, one
specifies an $\SO(1,3)$-connection $A$ together with a cotetrad field
$e$ rather than a metric tensor. Fixing $A|_{\del M}$ at the boundary,
one can derive first order field equations in the bulk which are
equivalent to the Einstein equations provided that the cotetrad is
non-degenerate. The theory is invariant under space-time
diffeomorphisms $f\colon M\to M$.

For the path integral, consider a smooth $4$-manifold $M$ whose
boundary $\del M=\Sigma_i\cup\Sigma_f$ (Figure~\ref{fig_manifolds}(a))
is the disjoint union of two smooth $3$-manifolds $\Sigma_i$ and
$\Sigma_f$ to which we associate Hilbert spaces $\mathcal{H}_j$ of
$3$-geometries, $j=i,f$. These contain suitable wave functionals of
connections $A|_{\Sigma_j}$. We denote the connection eigenstates by
$\ket{A|_{\Sigma_j}}$. The path integral,
\begin{equation}
\label{eq_pathint}
  \bra{A|_{\Sigma_f}}T_M\ket{A|_{\Sigma_i}} =
    \int_{A|_{\del M}}\mathcal{D}A\,\mathcal{D}e\,
      \exp\bigl(\frac{i}{\hbar}S\bigr),
\end{equation}
is the sum over all connections $A$ matching $A|_{\del M}$, and over
all $e$. It yields the matrix elements of a linear map
$T_M\colon\mathcal{H}_i\to\mathcal{H}_f$ between states of
$3$-geometry. For fixed boundary data, \eqref{eq_pathint} is
diffeomorphism invariant in the bulk. If $\Sigma_i\cong\Sigma_f$ are
diffeomorphic, we can identify $\Sigma=\Sigma_i=\Sigma_f$ and
$\mathcal{H}=\mathcal{H}_i=\mathcal{H}_f$ and consider $M^\prime$
(Figure~\ref{fig_manifolds}(b)). Provided that the trace over
$\mathcal{H}$ can be defined, the \emph{partition function},
\begin{equation}
\label{eq_partition}
  Z(M^\prime) := \tr_\mathcal{H}T_M = \int\mathcal{D}A\,\mathcal{D}e\,
      \exp\bigl(\frac{i}{\hbar}S\bigr),
\end{equation}
where the integral is now unrestricted, is a dimensionless number
which depends only on the diffeomorphism class of the smooth manifold
$M^\prime$~\cite{At88,Ba95,Cr95}.

Due to the potentially infinite-dimensional Hilbert space
$\mathcal{H}$, the partition function~\eqref{eq_partition} may be
formally divergent. In $2+1$ space-time dimensions, the following
examples are known: The Turaev--Viro model~\cite{TuVi92} (signature
$(1,1,1)$ and quantized positive cosmological constant) yields
finite-dimensional Hilbert spaces and a well-defined $Z(M^\prime)$
whereas the Ponzano--Regge model~\cite{PoRe68} (signature $(1,1,1)$
and zero cosmological constant) and the model with Lorentzian
signature $(-1,1,1)$~\cite{Fr01,Da01} have infinite-dimensional
Hilbert spaces and divergent $Z(M^\prime)$.

%
\section{Combinatorial triangulations}
%

We assume that we can first focus on the partition
function~\eqref{eq_partition} and then learn how to fix $A$ on
sub-manifolds in order to go back to~\eqref{eq_pathint} which is the
actual physically interesting object.

Any $(3+1)$-dimensional smooth space-time manifold $M$ is
characterized by its \emph{Whitehead triangulation}. This is a purely
combinatorial triangulation in terms of $k$-simplices, $0\leq k\leq 4$
(vertices, edges, triangles, tetrahedra, $4$-simplices), which
characterizes the topology and the differentiable structure of
$M$. Combinatorial here means that we specify which triangles are
contained in the boundary of which tetrahedra, \etc. Under well-known
conditions~\cite{RoSa72}, we can glue the simplices according to the
combinatorial data and obtain a $4$-manifold whose differentiable
structure is unique up to
diffeomorphism~\cite{HiMa74,Ku62}. Conversely, any two diffeomorphic
manifolds have equivalent (PL-isomorphic) triangulations~\cite{Wh40}.

If $M$ is compact with boundary $\del M$, the triangulation can be
chosen to consist of only a finite number of simplices. The generic
quantum geometry measurement involves such a compact $M$ and is given
by the generalization of the r.h.s.\ of~\eqref{eq_pathint} to generic
boundary~\cite{Oe03}. Let $M=[0,1]\times B^3$ where $B^3$ is the
$3$-ball. The boundary $\del M$ is the union of $\{0\}\times B^3$
(initial preparation of a space-like $3$-geometry), $[0,1]\times S^2$
(time-like geometry which represents a classical clock in the
laboratory surrounding the experiment while the system is kept
isolated), and $\{1\}\times B^3$ (final measurement of the space-like
$3$-geometry). Measurements of this type involve both classical
observers and a classical clock in the laboratory. They can in
principle be defined operationally.

Notice that some authors call other diffeomorphism invariant
expressions `observables', for example, expectation values of scalars
with respect to the measure of~\eqref{eq_partition}. We do not discuss
these global expressions since it is not known whether their
measurement can be defined operationally.

\begin{figure}[t]
\begin{center}
\input{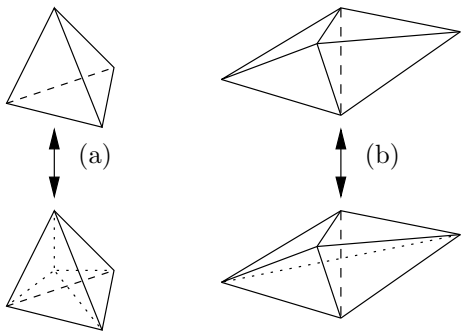}
\end{center}
\caption{
\label{fig_pachner}
  Pachner moves in $d=3$. (a) the $1\leftrightarrow 4$ move subdivides
  one tetrahedron into four. (b) the $2\leftrightarrow 3$ move changes
  the subdivision of a diamond from two tetrahedra to three ones
  (glued along the dotted line in the bottom picture).  }
\end{figure}

Let us sketch when triangulations are equivalent: two finite
triangulations are equivalent if and only if they are related by a
finite sequence of local modifications, called \emph{Pachner
moves}~\cite{Pa91}, which are elementary shellings for manifolds with
boundary and bistellar moves in the bulk. The latter change the
triangulation only inside some given polyhedron and replace one
possible subdivision of its interior by another
one. Figure~\ref{fig_pachner} shows the moves in $d=3$. The moves in
$d=4$ are illustrated in detail in~\cite{CaKa99}.

Sequences of Pachner moves are the combinatorial equivalent of
applying space-time diffeomorphisms. The partition
function~\eqref{eq_partition} and the path integral~\eqref{eq_pathint}
in the bulk therefore have to be invariant under Pachner moves. The
same holds, for example, for expectation values of scalars under the
measure of~\eqref{eq_partition}.

Having understood the moves, we can interpret the role of the
triangulation physically. Given the experience with lattice field
theory, the naive idea would be to think that the triangulation
provided a cut-off by introducing a physical length for the edges of
the triangulation. In view of the move of Figure~\ref{fig_pachner}(a)
which subdivides one simplex into four and which leaves all physical
predictions invariant, this cannot be true! The triangulation is
rather purely combinatorial. It does not carry any metric
information. The choice of one of several equivalent triangulations is
completely analogous to the choice of one of several atlases of
coordinate systems for the same smooth manifold. It has no physical
significance at all.

The metric data encoded in the path integral~\eqref{eq_pathint} and in
particular the expected Planck scale cut-off rather have a dynamical
origin. The known models in $2+1$ dimensions mentioned above,
dynamically assign metric information to the simplices of the
triangulation. The Turaev--Viro and the Ponzano--Regge model, for
example, endow the $1$-simplices (edges) with lengths $(j+1/2)\ell_P$
where $\ell_P=\hbar G/c^2$ denotes the Planck length and $j$ is a
half-integer. In $3+1$ dimensions, it is expected that the Planck area
$a_P=\hbar G/c^3$ plays the role of the fundamental geometric quantum.

%
\section{No local degrees of freedom}
%

\begin{figure}[t]
\begin{center}
\input{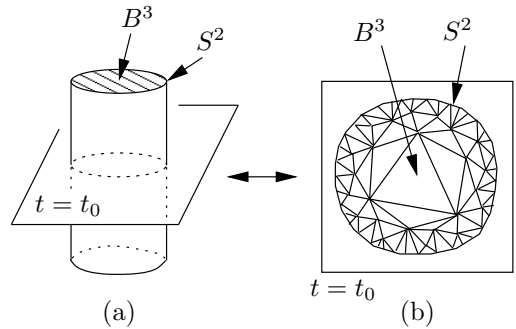}
\end{center}
\caption{
\label{fig_tube}
  (a) A tubular region of topology $\R\times B^3$ with boundary $\R\times
  S^2$ in standard space $\R^4$. (b) A slice at foliation parameter
  $t=t_0$. The triangulation of the boundary $S^2$ is as fine as
  required. In the interior of $B^3$, Pachner moves have been used to
  simplify the triangulation.
}
\end{figure}

As an application, let us consider a tubular space-time region
$M=\R\times B^3\subseteq\R^4$, Figure~\ref{fig_tube}(a), with boundary
$\del M=\R\times S^2$. Figure~\ref{fig_tube}(b) shows a slice of~(a)
for some foliation parameter $t=t_0$.

The experiment performed on the space-time region $M$ and measured
through $\del M$, requires a certain resolution which determines how
many simplices are needed in $\del M$. Beyond these boundary
conditions, we are free to choose any equivalent triangulation in the
bulk, in particular a very coarse one, without affecting any physical
prediction. Figure~\ref{fig_tube}(b) is deliberately drawn in a very
suggestive way. Of course, the apparent size of the simplices in the
Figure is completely arbitrary. What matters is only the
combinatorics.

Since the path integral dynamically assigns multiples of the Planck
length $\ell_P/2$ to the edges in $d=2+1$ or multiples of the Planck
area $a_P$ to the triangles in $d=3+1$, there is the following
indirect bound on the number of simplices in $\del M$ that is
compatible with a given classical geometry.

Whenever the quantum theory of gravity on $M$ has a classical limit in
which the area of $S^2$ is $A$, it makes sense to increase the number
$n$ of triangles in $S^2$ only up to $n\sim A/a_P$. Triangulations
with more triangles in $\del M$ are always assigned larger total areas
and their states therefore be suppressed in the very same classical
limit.

We have seen that the degrees of freedom are associated with the
simplices only up to equivalence of triangulations. In particular,
their number does not increase with the volume of the spatial region
$B^3$ (consider $t$ as `time'). The minimum number of simplices is
rather determined by the boundary conditions we impose on $S^2$, and
the number of simplices in the boundary is naturally bounded by
$A/a_P$ whenever there emerges a classical geometry such that $S^2$
has a total area of $A$.

Standard QFT, in contrast, essentially assigns harmonic oscillators,
each with an infinite-dimensional Hilbert space, to the uncountably
many points of space. We have reduced this to Hilbert spaces
associated with only a finite number of simplices. The major
conceptual difference is that standard QFT constructs Hilbert spaces
for the fields \emph{relative} to a fixed Riemannian $3$-manifold,
whereas the Hilbert spaces of~\eqref{eq_pathint} represent equivalence
classes of $3$-geometries up to spatial diffeomorphisms.

We note that none of our arguments changes substantially if we add
matter or gauge fields or if we formulate the classical theory in
terms of other variables. Recall that the symmetry under space-time
diffeomorphisms is not special to general relativity, but shared by
many other classical field theories, indicating a potential impact on
particle physics in general. Further details on path integral
quantization of general relativity and invariants of smooth manifolds
will be presented in~\cite{Pf04}.

\acknowledgments

I am grateful to Marco Mackaay and Louis Crane for advice on
$4$-manifolds. I would like to thank them and Lee Smolin for
stimulating discussions.


\begin{thebibliography}{999}
\bibitem{Ho93}
\textsc{G.~t'~Hooft}: Dimensional reduction in quantum gravity.
\newblock In \textsl{Salamfestschrift: a collection of talks}, eds. A.~Ali,
  J.~Ellis {\upshape and} S.~Randjbar-Daemi. World Scientific, Singapore, 1993,
  pp. 284--296, \texttt{gr-qc/9310026}.

\bibitem{Bo02}
\textsc{R.~Bousso}: The holographic principle.
\newblock \textsl{Rev.\ Mod.\ Phys.} \textbf{74} (2002) 825--874,
  \texttt{hep-th/0203101}.

\bibitem{Wa84}
\textsc{R.~M. Wald}: General relativity.
\newblock The University of Chicago Press, Chicago, 1984.

\bibitem{At88}
\textsc{M.~F. Atiyah}: Topological quantum field theories.
\newblock \textsl{IHES Publ.\ Math.} \textbf{68} (1988) 175--186.

\bibitem{Ba95}
\textsc{J.~W. Barrett}: Quantum gravity as a topological quantum field theory.
\newblock \textsl{J.\ Math.\ Phys.} \textbf{36}, No.~11 (1995) 6161--6179,
  \texttt{gr-qc/9506070}.

\bibitem{Cr95}
\textsc{L.~Crane}: Clock and category: Is quantum gravity algebraic?
\newblock \textsl{J.\ Math.\ Phys.} \textbf{36}, No.~11 (1995) 6180--6193,
  \texttt{gr-qc/9504038}.

\bibitem{TuVi92}
\textsc{V.~G. Turaev {\upshape and} O.~Y. Viro}: State sum invariants of
  $3$-manifolds and quantum $6j$-symbols.
\newblock \textsl{Topology} \textbf{31}, No.~4 (1992)
  865--902.

\bibitem{PoRe68}
\textsc{G.~Ponzano {\upshape and} T.~Regge}: Semiclassical limit of Racah
  coefficients.
\newblock In \textsl{Spectroscopic and group theoretical methods in physics},
  ed. F.~Bloch. North Holland Publications, Amsterdam, 1968, pp. 1--58.

\bibitem{Fr01}
\textsc{L.~Freidel}: A Ponzano--Regge model of Lorentzian 3-dimensional
  gravity.
\newblock \textsl{Nucl.\ Phys.\ Proc.\ Suppl.} \textbf{B 88} (2000) 237--240,
  \texttt{gr-qc/0102098}.

\bibitem{Da01}
\textsc{S.~Davids}: A state sum model for $(2+1)$ Lorentzian quantum gravity.
\newblock {PhD} thesis, University of Nottingham (2001).
\newblock Preprint \texttt{gr-qc/0110114}.

\bibitem{RoSa72}
\textsc{C.~P. Rourke {\upshape and} B.~J. Sanderson}: Introduction to
  piecewise-linear topology.
\newblock Ergebnisse der Mathematik und ihrer Grenzgebiete 69. Springer,
  Berlin, 1972.

\bibitem{HiMa74}
\textsc{M.~W. Hirsch {\upshape and} B.~Mazur}: Smoothings of piecewise linear
  manifolds.
\newblock Annals of Mathematics Studies 80. Princeton University Press,
  Princeton, 1974.

\bibitem{Ku62}
\textsc{N.~H. Kuiper}: On `La nullit{\'e} de $\pi_0(\mathrm{Diff} S^3)$' by
  J.~Cerf.
\newblock Mathematical Reviews No.~6641 a--d Vol.~33 (1963) 1126.

\bibitem{Wh40}
\textsc{J.~H.~C. Whitehead}: On $C^1$-complexes.
\newblock \textsl{Ann.\ Math.} \textbf{41}, No.~4 (1940) 809--824.

\bibitem{Oe03}
\textsc{R.~Oeckl}: Schr{\"o}dinger's cat and the clock: Lessons for quantum
  gravity.
\newblock \textsl{Class.\ Quant.\ Grav.} \textbf{20} (2003) 5371--5380,
  \texttt{gr-qc/0306007}.

\bibitem{Pa91}
\textsc{U.~Pachner}: PL homeomorphic manifolds are equivalent by elementary
  shellings.
\newblock \textsl{Eur.\ J.\ Comb.} \textbf{12} (1991) 129--145.

\bibitem{CaKa99}
\textsc{J.~S. Carter, L.~H. Kauffman {\upshape and} M.~Saito}: Structures and
  diagrammatics of four dimensional topological lattice field theories.
\newblock \textsl{Adv.\ Math.} \textbf{146} (1999) 39--100,
  \texttt{math.GT/9806023}.

\bibitem{Pf04}
\textsc{H.~Pfeiffer}: Quantum general relativity and the classification of
  smooth manifolds (2004).
\newblock Preprint \texttt{gr-qc/0404088}.

\end{thebibliography}
\end{document}